\documentclass[ALICE,manyauthors]{cernphprep}
\pdfoutput=1
\usepackage{hyperref}
\usepackage{lineno}

\usepackage[english]{babel}
\usepackage{graphicx}
\usepackage{verbatim}
\usepackage{amsfonts}
\usepackage{makeidx}
\usepackage{color}
\usepackage{amsmath}
\usepackage{lineno}
\usepackage{epsfig}
\usepackage{epstopdf}
\usepackage{hyperref}
\usepackage{cite}
\usepackage{textcomp}
\usepackage{multirow}
\newcommand{\sqrtsNN}{\sqrt{s_{\rm \scriptscriptstyle NN}}}

\newcommand{\GeV}{\mathrm{GeV}}
\newcommand{\TeV}{\mathrm{TeV}}

\newcommand{\mum}{\mathrm{\mu m}}

\newcommand{\mub}{\mathrm{\mu b}}

\newcommand{\PbPb}{\mbox{Pb--Pb}}

\newcommand{\Raa}{R_{\rm AA}}

\newcommand{\pt}{p_{\rm T}}

\newcommand{\DtoKpi}{{\rm D}^0 \to {\rm K}^-\pi^+}
\newcommand{\DtoKpipi}{{\rm D}^+\to {\rm K}^-\pi^+\pi^+}

\newcommand{\DstartoDpitoKpipi}{{\rm D}^{*+} \to {\rm D}^0 \pi^+ \to {\rm K}^{-} \pi^+ \pi^+}

\newcommand{\Dzero}{{\rm D^0}}

\newcommand{\Dstar}{{\rm D^{*+}}}

\newcommand{\Dplus}{{\rm D^+}}

\newcommand{\qTPC}{q_2^{\rm TPC}}
\newcommand{\qVZEROA}{q_2^{\rm V0A}}

\begin{document}%

\begin{titlepage}
\PHyear{2018}
\PHnumber{260}      
\PHdate{23 September}  
%

\title{Event-shape engineering for the D-meson elliptic flow in mid-central Pb--Pb collisions at $\pmb{\sqrtsNN = 5.02}$~TeV}
\ShortTitle{ESE for the D-meson $v_2$ in Pb--Pb collisions at $\sqrtsNN=5.02$~TeV}   

\Collaboration{ALICE Collaboration\thanks{See Appendix~\ref{app:collab} for the list of collaboration members}}
\ShortAuthor{ALICE Collaboration} 

\begin{abstract}
The production yield of prompt D mesons and their elliptic flow coefficient $v_2$ were measured with the Event-Shape Engineering (ESE) technique applied to mid-central (10--30\% and 30--50\% centrality classes) Pb--Pb collisions at the centre-of-mass energy per nucleon pair $\sqrtsNN=5.02$~TeV, with the ALICE detector at the LHC. The ESE technique allows the classification of events, belonging to the same centrality, according to the azimuthal anisotropy of soft particle production in the collision. The reported measurements give the opportunity to investigate the dynamics of charm quarks in the Quark--Gluon Plasma and provide information on their participation in the collective expansion of the medium. D mesons were reconstructed via their hadronic decays at mid-rapidity, $|\eta|<0.8$, in the transverse momentum interval $1<\pt<24$~GeV/$c$. The $v_2$ coefficient is found to be sensitive to the event-shape selection confirming a correlation between the D-meson azimuthal anisotropy and the collective expansion of the bulk matter, while the per-event D-meson yields do not show any significant modification within the current uncertainties.
\end{abstract}
\end{titlepage}
\setcounter{page}{2}

\section{Introduction}
Quantum Chromo-Dynamics (QCD) calculations on the lattice predict the existence of a plasma of deconfined quarks and gluons, known as the Quark--Gluon Plasma (QGP)~\cite{Karsch:2006xs,Borsanyi:2010bp,Bazavov:2011nk,Florkowski:2015rxg}. The transition from the hadronic phase to the QGP state occurs at high temperatures and energy densities, which can be reached in collisions of heavy nuclei at ultra-relativistic energies. The QGP created in ultra-relativistic heavy-ion collisions was found to behave as a nearly ideal fluid (i.e. with a small shear viscosity over entropy density ratio, $\eta/s$), undergoing an expansion that can be described by relativistic hydrodynamics~\cite{Song:2010mg,Qiu:2011hf,Gale:2012rq,Adams:2004bi,Alver:2006wh,ALICE:2011ab}. 

Heavy flavours (charm and beauty quarks), due to their large masses, $m_{\rm c}\approx 1.3$~GeV/$c^2$ and $m_{\rm b}\approx 4.5$~GeV/$c^2$, are predominantly produced in hard-scattering processes characterised by timescales shorter than the QGP formation time~\cite{Liu:2012ax,BraunMunzinger:2007tn,Andronic:2015wma,Aarts:2016hap}. Thus, they experience the entire evolution of the medium interacting with its constituents via inelastic (gluon radiation)~\cite{Gyulassy:1990ye,Baier:1996sk,Prino:2016cni} and elastic (collisional)~\cite{Braaten:1991we} QCD processes. 
Such interactions with the medium constituents can also lead to a modification of the hadronisation mechanism with respect to the fragmentation in vacuum: a significant fraction of low- and intermediate-momentum charm and beauty quarks can hadronise via recombination with other quarks from the medium~\cite{Greco:2003mm,Andronic:2003zv,Plumari:2017ntm}.

Heavy-flavour hadrons are effective probes of the properties of the medium produced in heavy-ion collisions. A strong modification of their transverse-momentum ($p_{\rm T}$) distributions in heavy-ion collisions with respect to pp collisions was observed at RHIC~\cite{Adare:2010de,Abelev:2006db,Adamczyk:2014uip,Adler:2005xv} and LHC energies~\cite{Adam:2015sza,Abelev:2012qh,Adam:2016khe,Adam:2016wyz,Khachatryan:2016ypw,Sirunyan:2017xss,Acharya:2018hre}. 
In particular, the observed suppression of the yield of heavy-flavour hadrons in central nucleus--nucleus collisions relative to pp collisions scaled by the number of nucleon--nucleon collisions provides compelling evidence of the heavy-quark energy loss in deconfined strongly-interacting matter~\cite{Andronic:2015wma,Prino:2016cni}. 

Further insight into the interactions of heavy quarks with the medium can be obtained through measurements of the azimuthal distributions of heavy-flavour hadrons in heavy-ion collisions. The initial spatial anisotropy present in the early stages of nucleus--nucleus collisions is converted via multiple interactions into an azimuthally anisotropic distribution in momentum space of the produced particles~\cite{Ollitrault:1992bk,Voloshin:1994mz}. This anisotropy can be characterised in terms of the Fourier coefficients $v_n$ of the azimuthal distribution of particle momenta relative to the symmetry-plane angles $\Psi_n$ (for the $n^{\rm th}$ harmonic)~\cite{Voloshin:1994mz,Poskanzer:1998yz}. The values of the Fourier coefficients depend on the geometry of the collision, the fluctuations in the distributions of nucleons within the nuclei\cite{Qin:2010pf}, and the dynamics of the expansion. The second-order coefficient $v_2=\langle \cos [2(\varphi-\Psi_2)]\rangle$, where $\varphi$ is the particle momentum azimuthal angle and the brackets indicate the average over all the measured particles in the considered events, is usually denoted as elliptic flow. In non-central heavy-ion collisions, it represents the dominant term in the Fourier expansion~\cite{Ollitrault:1992bk,Poskanzer:1998yz}. 
The measurement of the azimuthal anisotropy of heavy-flavour hadrons at low $p_{\rm T}$ is sensitive to whether charm quarks take part in the collective expansion of the medium~\cite{Batsouli:2002qf}, as well as to the fraction of heavy-flavour hadrons hadronising via recombination with flowing light quarks~\cite{Molnar:2004ph,Greco:2003vf}.
At high $p_{\rm T}$, it can constrain the path-length dependence of heavy-quark in-medium energy loss~\cite{Gyulassy:2000gk,Shuryak:2001me}.
A positive $v_2$ in the heavy-flavour sector was observed at RHIC in Au--Au collisions at a centre-of-mass energy per nucleon pair $\sqrt{s_{\rm NN}}=200$~GeV~\cite{Adare:2010de,Adamczyk:2014yew,Adamczyk:2017xur} and at the LHC in Pb--Pb collisions at $\sqrt{s_{\rm NN}}=2.76$~TeV~\cite{Abelev:2013lca,Abelev:2014ipa,Adam:2016ssk,Adam:2015pga}. Evidence of a positive D-meson $v_2$ was also reported in Pb--Pb collisions at $\sqrt{s_{\rm NN}}=5.02$~TeV by the ALICE~\cite{Acharya:2017qps} and CMS~\cite{Sirunyan:2017plt} Collaborations. The anisotropic flow of beauty quarks was investigated by the CMS Collaboration through the measurement of non-prompt J/$\psi$ elliptic flow~\cite{Khachatryan:2016ypw}. The D-meson results are described by theoretical calculations including mechanisms that impart a positive $v_2$ to charm quarks through the interactions with the hydrodynamically-expanding medium, namely collisional processes, and recombination of charm and light quarks~\cite{Uphoff:2012gb,He:2014cla,Monteno:2011gq,Djordjevic:2015hra,Cao:2013ita,Song:2015ykw,Nahrgang:2013xaa,Uphoff:2014hza,Beraudo:2014boa,Cao:2017hhk}. According to these model calculations, the same mechanisms affect the beauty-quark propagation in the medium, although the beauty-hadron $v_2$ is expected to be smaller than that of charm hadrons and to have a different transverse momentum dependence due to the large mass of the b quarks.   
Precise measurements of $v_2$ of heavy-flavour hadrons help to constrain model parameters, e.g., the heavy-quark spatial diffusion coefficient $D_s$ in the QGP, which is related to the relaxation time (or the time scale for equilibration) of the heavy quarks inside the QGP~\cite{Moore:2004tg,Acharya:2017qps}.

The Event Shape Engineering (ESE) technique~\cite{Schukraft:2012ah} can be used to further investigate the dynamics of heavy quarks in the medium. This technique has already been exploited in the light-flavour sector to study the interplay between the initial geometry of the nucleus--nucleus collisions and the subsequent evolution of the system~\cite{Adam:2015eta,Aad:2015lwa}, and to investigate the Chiral Magnetic Effect (CME)~\cite{Acharya:2017fau,Sirunyan:2017quh}. The ESE technique is based on the observation of a large event-by-event $v_n$ variation at fixed collision centrality~\cite{Abelev:2012di}. Hydrodynamic calculations show a linear correlation between the final state $v_2$ (and $v_3$) and the corresponding eccentricities in the initial state $\epsilon_2$ (and $\epsilon_3$) for small values of $\eta/s$~\cite{Heinz:2013th,Gardim:2012dc,Voloshin:2008dg}. 
These observations suggest the possibility to select heavy-ion collisions with different initial geometrical shape on the basis of the magnitude of the average bulk flow.

The ESE technique provides a tool to investigate the correlation between the flow coefficients of D mesons and soft hadrons: measuring the D-meson $v_2$ in classes of events in a given centrality interval, but with different magnitude of the average event flow can be useful to study the interplay between the anisotropic flow of heavy quarks and that of the bulk matter. In addition, it could provide insights on how the fluctuations in the initial geometry of the system affect the path-length-dependent energy loss experienced by the heavy quarks in the QGP. For these reasons, the application of the ESE technique to the D-meson $v_2$ measurements could be exploited to infer more information on the dynamics of the charm quark in the QGP and has the potential to set additional constraints on parameters of model calculations implementing heavy-quark transport in an hydrodynamically expanding medium~\cite{Ke:2018tsh}.
Model calculations for the correlation between the $v_2$ values of soft hadrons and heavy-flavour mesons on an event-by-event basis have recently become available. 
A linear correlation between the high-$\pt$ D-meson $v_2$, which originates from the path-length dependence of in-medium energy loss, and the elliptic flow of charged hadrons is predicted in~\cite{Prado:2016szr}, based on a model for charm-quark energy loss in a medium described event-by-event with viscous hydrodynamics. Within the heavy-quark transport model of~\cite{Gossiaux:2017nwz}, an almost linear correlation is obtained between the $v_2$ of pions and that of $\Dzero$ mesons with $\pt>2~\GeV/c$, which is dominated by low-$\pt$ mesons and is therefore sensitive to the degree of thermalisation of charm quarks with the collectively expanding medium.
According to these calculations, the initial system ellipticity is converted into parton flow with a similar efficiency for bulk and charm quarks, despite the different production mechanisms, dynamics and hadronisation of heavy quarks and light partons forming the bulk of the medium. 

Finally, the measurement of the D-meson yields at low and intermediate $\pt$ in ESE-selected events allows the investigation of a possible interplay between elliptic and radial flow, already observed for charged and identified particles~\cite{Adam:2015eta}. This observation is possibly related to the correlation between the density of participant nucleons and the initial eccentricity of the collision. For high-$\pt$ D mesons, the measurement of the yields in collisions with different initial eccentricity via the ESE technique could further constrain in-medium energy loss models.

In this paper, the $\Dzero$, $\Dplus$ and $\Dstar$ meson $v_2$ in Pb--Pb collisions at $\sqrt{s_{\rm NN}}=5.02$~TeV in the 10--30\% centrality class are presented and compared to the results in the 30--50\% centrality class published in~\cite{Acharya:2017qps}. The $\Dzero$ and $\Dplus$ $v_2$ obtained with ESE and the measurement of D-meson yield ratios in ESE-selected events in the 10--30$\%$ and 30--50$\%$ centrality classes are reported as well.

\section{Data analysis}
\label{sec:analysis}
The $\Dzero$, $\Dplus$ and $\Dstar$ mesons were reconstructed at mid-rapidity, exploiting the tracking and particle identification capabilities of the ALICE detector at the LHC. A detailed description of the ALICE experimental apparatus and its performance can be found in~\cite{Aamodt:2008zz,Abelev:2014ffa}. The main detectors used for the analysis presented in this paper are the Inner Tracking System (ITS), a six-layer silicon detector used to track charged particles and for the reconstruction of primary and secondary vertices; the Time Projection Chamber (TPC), which provides track reconstruction as well as particle identification via the measurement of the specific ionisation energy loss ${\rm d}E/{\rm d}x$; the Time-Of-Flight (TOF) detector, an array of Multi-Gap Resistive Plate Chambers that provides particle identification via the measurement of the flight time of the particles. These detectors cover the pseudorapidity interval $|\eta|<0.9$ and are located in a large solenoidal magnet providing a uniform magnetic field of 0.5 T parallel to the LHC beam direction. In addition, two detectors were used for the event selection and classification: the V0 detector, which consists of two arrays of 32 scintillators each, covering the full azimuth in the pseudorapidity intervals $-3.7 < \eta < -1.7$ (V0C) and $2.8 < \eta < 5.1$ (V0A); and the Zero Degree Calorimeters (ZDC), located at $112.5$ m from the interaction point on either side, to detect spectator neutrons and protons of the colliding nuclei. 

	The analysed data sample consists of $\PbPb$ collisions at $\sqrtsNN=5.02~\TeV$ collected using a minimum-bias interaction trigger that required coincident signals in both scintillator arrays of the V0 detector. Events were selected offline by using the V0 and the neutron ZDC timing information, to remove contaminations produced by the interaction of the beams with residual gas in the vacuum pipe. Only events with a reconstructed primary vertex within $\pm10$ cm from the centre of the detector along the beam line were analysed. Events satisfying the aforementioned selections were divided in centrality classes, defined in terms of percentiles of the hadronic Pb--Pb cross section. This classification was based on a fit to the sum of the signal amplitudes measured in the V0 detectors. The fit function assumes the Glauber model~\cite{Glauber:1970jm,Miller:2007ri} combined with a two-component model for particle production~\cite{Abelev:2013qoq}. The number of events in each centrality class considered for this analysis (10--30\% and 30--50\%) is about $20.7\times 10^6$, corresponding to an integrated luminosity of about $13~\mub^{-1}$. The events in each centrality class were further divided in samples with different average elliptic anisotropy of final-state particles, selected according to the magnitude of the second-order harmonic reduced flow vector $q_2$~\cite{Adler:2002pu,Voloshin:2008dg}, defined as 
		\begin{equation}
		q_2 = |\pmb{Q}_2|/\sqrt{M},	
		\label{eq:q2}
		\end{equation}
		where $M$ is the multiplicity (number of tracks used in the $q_2$ calculation) and 	
		\begin{equation}
		\pmb{Q}_2 = \left(
		\begin{array}{c}
			\sum_{i=1}^{M}\cos(2\varphi_i)\\
			\sum_{i=1}^{M}\sin(2\varphi_i)
		\end{array}
		\right)
		\end{equation}
		is the second-order flow vector, which is built starting from the azimuthal angles ($\varphi_i$) of the considered particles. The denominator in Eq.~\ref{eq:q2} is introduced to remove the dependence of $|\pmb{Q}_2|$ on $\sqrt{M}$ in the absence of flow~\cite{Voloshin:2008dg}. 
		
		The $\pmb{Q}_2$ vector was measured using charged tracks reconstructed in the TPC ($\qTPC$), with $|\eta|<0.8$ and $0.2<\pt<5~\GeV/c$, to exploit the good $\varphi$ resolution of the TPC and the large multiplicity at midrapidity, which are crucial to maximise the selectivity of $q_2$. In order to remove autocorrelations between D mesons and $q_2$, the tracks used to form the D-meson candidates were excluded from the computation of $q_2$. However, with this definition of $q_2$, some residual non-flow correlations (i.e. correlations among particle emission angles not induced by the collective expansion but rather by particle decays and jet production) could still be included. As shown in ~\cite{Adam:2015eta}, the introduction of a pseudorapidity separation of more than one unit between the region used to calculate $q_2$ and the region used to measure the observables would suppress unwanted non-flow contributions. Therefore, to investigate a possible effect induced by non-flow contaminations, $q_2$ was also measured using the V0A detector ($\qVZEROA$), allowing for a pseudorapidity separation of at least 2 units between the D-meson decay tracks and the particles used for the $q_2$ determination. In this case, the $\pmb{Q}_2$ vector was calculated from the azimuthal distribution of the energy deposition measured in the V0A detector, and its components are given by
			\begin{equation}
			Q_{2,x} = \sum_{i=1}^{N_{\rm sectors}} w_i\cos (2\varphi_i),\quad Q_{2,y} = \sum_{i=1}^{N_{\rm sectors}} w_i\sin (2\varphi_i),
			\label{eq:QvecV0}
			\end{equation}
			where the sum runs over the 32 sectors ($N_{\rm sectors}$) of the V0A detector, $\varphi_i$ is the angle of the centre of the sector $i$ and $w_i$ is the amplitude measured in sector $i$, once the gain equalisation method~\cite{Selyuzhenkov:2007zi} is applied to correct effects of non-uniform acceptance.
		The comparison between the two ESE selections is discussed in Sec.~\ref{subsec:nonflw_q2selectivity}.
				
\begin{figure*}[!t]
\begin{center}
\includegraphics[width=.9\textwidth]{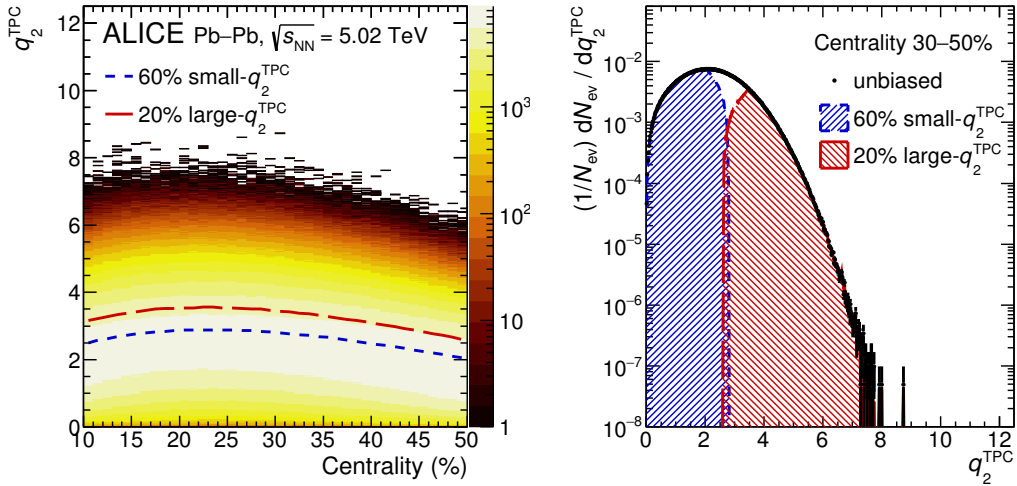}
\caption{Left: distribution of $\qTPC$ (see text for details) as a function of centrality in $\PbPb$ collisions at $\sqrtsNN=5.02~\TeV$. The red long-dashed and the blue short-dashed lines represent the limits for the 20\% and the 60\% of events with largest and smallest $\qTPC$, respectively. Right: $\qTPC$ distributions for the unbiased, small-$\qTPC$ and large-$\qTPC$ samples for the 30--50\% centrality class (see text for details).}
\label{fig:q2_vs_cent} 
\end{center}
\end{figure*}

		The left panel of Fig.~\ref{fig:q2_vs_cent} shows the centrality dependence of the $\qTPC$ distribution.
			As expected in case of large initial-state fluctuations, the $q_2$ distribution is broad and reaches values larger than twice the mean value~\cite{Schukraft:2012ah}. Moreover, because of the different average elliptic flow and multiplicity, the $q_2$ distribution changes as a function of centrality. Hence, a selection on a fixed value of $q_2$ would induce a non-flat centrality distribution, that would spoil the event-shape selection. For this reason, the selection of the events according to their $q_2$ was performed by defining $q_2$ percentiles in 1\%-wide centrality intervals. The results presented in the following sections are obtained in two ESE-selected classes, corresponding to the 60\% and the 20\% of events with smallest and largest $q_2$, respectively. The $\qTPC$ distributions for these classes in the 30--50\% centrality interval are displayed in the right panel of Fig.~\ref{fig:q2_vs_cent}. In the following, we will refer to these two classes as "small-$q_2$" and "large-$q_2$". In case of no event-shape selection, we will use the "unbiased" term.  

The D mesons, together with their charge conjugates, were reconstructed via their hadronic decay channels $\DtoKpi$, $\DtoKpipi$ and $\DstartoDpitoKpipi$. The D-meson candidates were built combining pairs and triplets of tracks with proper charge sign, $|\eta|<0.8$, $\pt>0.4~\GeV/c$, a minimum number of 70 (out of 159) associated space points in the TPC and no less than two hits (out of six) in the ITS, with at least one in the two innermost layers. For the soft pion produced in the $\Dstar$ decay, also tracks reconstructed only in the ITS, with at least three associated hits and with $\pt>0.1~\GeV/c$, were considered. These selections limit the D-meson rapidity acceptance, which drops steeply to zero for $|y|>0.6$ for $\pt=1~\GeV/c$ and $|y|>0.8$ for $\pt>5~\GeV/c$. Therefore, a $\pt$-dependent fiducial acceptance selection, $|y_{\rm D}| < y_{\rm fid}(\pt)$, was applied. The selection value, $y_{\rm fid}(\pt)$, was defined according to a second-order polynomial function, increasing from 0.6 to 0.8 in the range $1 < \pt < 5$~GeV/$c$, and fixed to a constant value of 0.8 for $\pt > 5~\GeV/c$. 

The D-meson candidate selection strategy for the reduction of the combinatorial background is similar to the one used in previous analyses~\cite{Abelev:2014ipa, Acharya:2017qps}. The selection of the $\Dzero$ and $\Dplus$ decay topology was based on the reconstruction of secondary vertices with a separation of a few hundred microns from the primary vertex ($c\tau\simeq 123$ and  $312~\mum$ for $\Dzero$ and $\Dplus$, respectively~\cite{Olive:2016xmw}). The main variables used to enhance the statistical significance and the signal-to-background ratio are the displacement of the decay tracks from the interaction vertex, the separation between the secondary and primary vertices and the pointing angle of the reconstructed D-meson momentum to the primary vertex. In the case of the strong decay of the $\Dstar$ meson, the secondary vertex cannot be resolved from the primary vertex, and therefore geometrical selections were applied on the displaced decay-vertex topology of the produced $\Dzero$ mesons. In addition, for $\Dzero$ and $\Dplus$ mesons, the normalised difference between the measured and expected transverse-plane impact parameters of each of the decay particles and the transverse-plane impact parameter to the primary vertex ($d_0^{xy}$) of the $\Dplus$-meson candidates were applied to suppress the fraction of D mesons coming from beauty-hadron decays (denoted by "feed-down" in the following) and hence reduce the associated systematic uncertainty. These selections were found to be especially effective for $\Dplus$ mesons, for which a rejection of the feed-down contribution up to 50\% at high $\pt$ was achieved.
The selection criteria for each D-meson species were optimised as a function of $\pt$ independently for the two centrality classes, because of the different combinatorial background. Within a given centrality class, the same selection criteria were applied in the different ESE-selected samples. In order to further reduce the combinatorial background, a particle identification for charged pions and kaons with the TPC and TOF detectors was applied, using a selection in units of resolution (at $\pm 3~\sigma$) around the expected mean values of ${\rm d}E/{\rm d}x$ and time of flight, respectively. 

	Monte Carlo simulations with a detailed description of the detector and its response, based on the GEANT3 transport package~\cite{Brun:1994aa}, were used to study the signal invariant-mass distributions and the reconstruction efficiencies, as described in the following. In the Monte Carlo simulation, the underlying Pb--Pb events at $\sqrtsNN = 5.02$~TeV were simulated using the HIJING v1.383 generator~\cite{Wang:1991hta} and $\rm c\overline c$ or $\rm b\overline b$ pairs were added with the PYTHIA v6.421 generator~\cite{Sjostrand:2006za} with Perugia-2011 tune~\cite{Skands:2010ak}. The generated D mesons were forced to decay into the hadronic channels of interest for the analysis.
		\begin{table}[!t]
		\centering	
		\renewcommand*{\arraystretch}{1.1}
		\begin{tabular}{cccc}
		\hline
		Centrality class & Detector for $\psi_2$ & Event-shape class & Event-plane resolution $R_2$\\
		\hline
		\multirow{6}{*}{10--30\%} &\multirow{3}{*}{V0} & unbiased &$0.8223 \pm 0.0001$  \\
		& & small-$\qTPC$ & $0.7809\pm0.0001$ \\
		& & large-$\qTPC$ & $0.9058\pm0.0001$ \\
		\cline{2-4}
		& \multirow{3}{*}{V0C}& unbiased &$0.7669 \pm 0.0001$  \\
		& & small-$\qVZEROA$ & $0.7390\pm0.0001$ \\
		& & large-$\qVZEROA$ & $0.8223\pm0.0001$ \\
		\hline
		\multirow{6}{*}{30--50\%} &\multirow{3}{*}{V0} & unbiased & $0.7708\pm0.0001$   \\
		& & small-$\qTPC$ & $0.7301 \pm 0.0001$ \\
		& & large-$\qTPC$ & $0.8646 \pm 0.0001$ \\
		\cline{2-4}
		& \multirow{3}{*}{V0C}& unbiased &$0.7077 \pm 0.0001$  \\
		& & small-$\qVZEROA$ & $0.6822\pm0.0001$ \\
		& & large-$\qVZEROA$ & $0.7597\pm0.0001$ \\
		\hline		
		\end{tabular}	
		\caption{Event-plane resolution $R_2$ in the 10--30\% and 30--50\% centrality classes for the unbiased, small-$q_2$ and large-$q_2$ samples. The quoted uncertainty is statistical only.}
		\label{tab:EPresol}
		\end{table}
	
	The D-meson elliptic flow, $v_2$, is measured using the Event-Plane (EP) method~\cite{Poskanzer:1998yz}. This analysis technique relies on the event-by-event estimate of the second-order harmonic symmetry plane $\Psi_2$ using the so-called event-plane angle
		\begin{equation}
			\psi_2 = \frac{1}{2}\tan^{-1}\bigg{(}\frac{Q_{2,y}}{Q_{2,x}}\bigg{)}.
		\end{equation}
	For the measurements of $v_2$ in the unbiased and $\qTPC$-selected samples, $\pmb{Q}_2$ was estimated with the full V0 detector using  Eq.~\ref{eq:QvecV0} (with $N_{\rm sectors}$ corresponding to the 64 sectors of the full V0 detector). In case of the ESE selection based on $\qVZEROA$, only the 32 sectors of the V0C were used for the $\psi_2$ determination, to avoid autocorrelations with the $q_2$ measurement. 

	After the topological and kinematical selections, the D-meson candidates were divided in two samples, according to their azimuthal angle relative to the event-plane angle $\Delta\varphi=\varphi_{\rm D} - \psi_2$, namely in-plane ($] -\frac{\pi}{4}, \frac{\pi}{4}]$ and $] \frac{3\pi}{4}, \frac{5\pi}{4}]$) and out-of-plane ($] \frac{\pi}{4}, \frac{3\pi}{4}]$ and $] \frac{5\pi}{4}, \frac{7\pi}{4}]$). The separation of at least 0.9 units of pseudorapidity ($|\Delta\eta|>0.9$) between the D mesons and the particles used to measure $\psi_2$, naturally ensured by the selection of D-meson decay tracks and the V0 (V0C) acceptance, suppresses non-flow contributions. The $v_2$ can therefore be expressed by the following equation~\cite{Abelev:2014ipa}
		\begin{equation}
			v_2\{{\rm EP}\} = \frac{1}{R_2}\frac{\pi}{4}\frac{N_{\rm in\textnormal{-}plane}-N_{\rm out\textnormal{-}of\textnormal{-}plane}}{N_{\rm in\textnormal{-}plane}+N_{\rm out\textnormal{-}of\textnormal{-}plane}},
		\label{eq:v2EP}
		\end{equation}	
		where $N_{\rm in\textnormal{-}plane}$ and $N_{\rm out\textnormal{-}of\textnormal{-}plane}$ are the D-meson raw yields in the two $\Delta\varphi$ intervals. The raw yields can be directly used in Eq.~\ref{eq:v2EP}, without an efficiency correction, since simulations showed that the D-meson reconstruction and selection efficiencies do not depend on $\Delta\varphi$~\cite{Abelev:2014ipa}. The factor $1/R_2$ in  Eq.~\ref{eq:v2EP} is the correction due to the finite resolution of the estimated $\psi_2$ angle. In case of $v_2$ measurements in the unbiased and $\qTPC$-selected samples, the event-plane resolution $R_2$ was determined by correlating three sub-events of charged particles reconstructed in the V0 itself, in the positive ($0<\eta<0.8$) and negative ($-0.8<\eta<0$) semivolumes of the TPC~\cite{Poskanzer:1998yz}. In case of ESE selection based on $\qVZEROA$, the three sub-events considered were the charged particles reconstructed in the V0C, in the V0A, and in the full volume of the TPC ($|\eta|<0.8$). The values of $R_2$ estimated in the 10--30\% and 30--50\% centrality classes, for the unbiased, small-$q_2$ and large-$q_2$ samples are reported in Tab.~\ref{tab:EPresol}. The $R_2$ factor is higher (lower) in the large-$q_2$ (small-$q_2$) class with respect to that evaluated for the unbiased sample and similarly in the V0 case than in the V0C one, since the event-plane resolution $R_2$ increases with increasing $v_2\sqrt{M}$~\cite{Poskanzer:1998yz}. 
			\begin{figure*}[!t]
			\begin{center}
			\includegraphics[width=1.\textwidth]{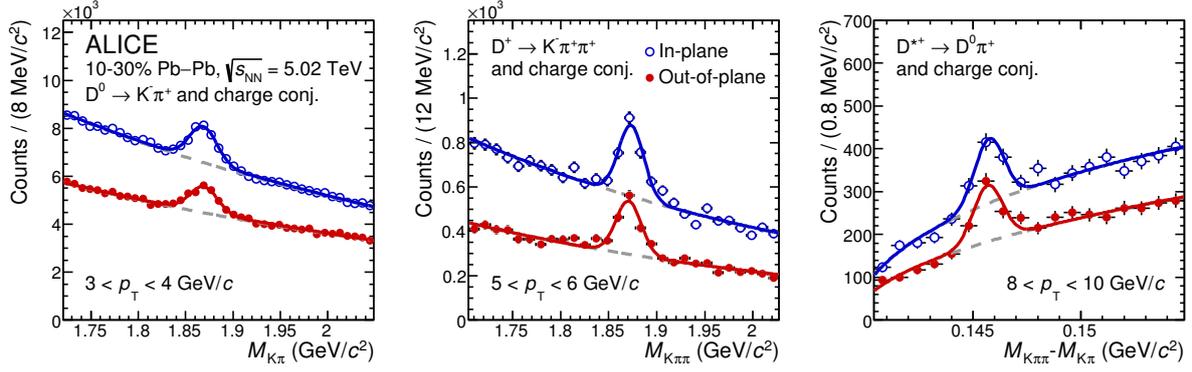}
			\caption{Invariant-mass distributions of $\Dzero$ candidates (left panel), $\Dplus$ candidates (middle panel) and mass-difference for $\Dstar$ candidates (right panel) in three $\pt$ intervals for the two $\Delta\varphi$ regions used in the EP method for $\PbPb$ collisions in the 10--30\% centrality class at $\sqrtsNN=5.02~\TeV$. The solid curves represent the total fit functions and the dotted curves the background functions, as described in the text.}
			\label{fig:massplots_unb} 
			\end{center}
			\end{figure*}

\begin{figure*}[!t]
			\begin{center}
			\includegraphics[width=.8\textwidth]{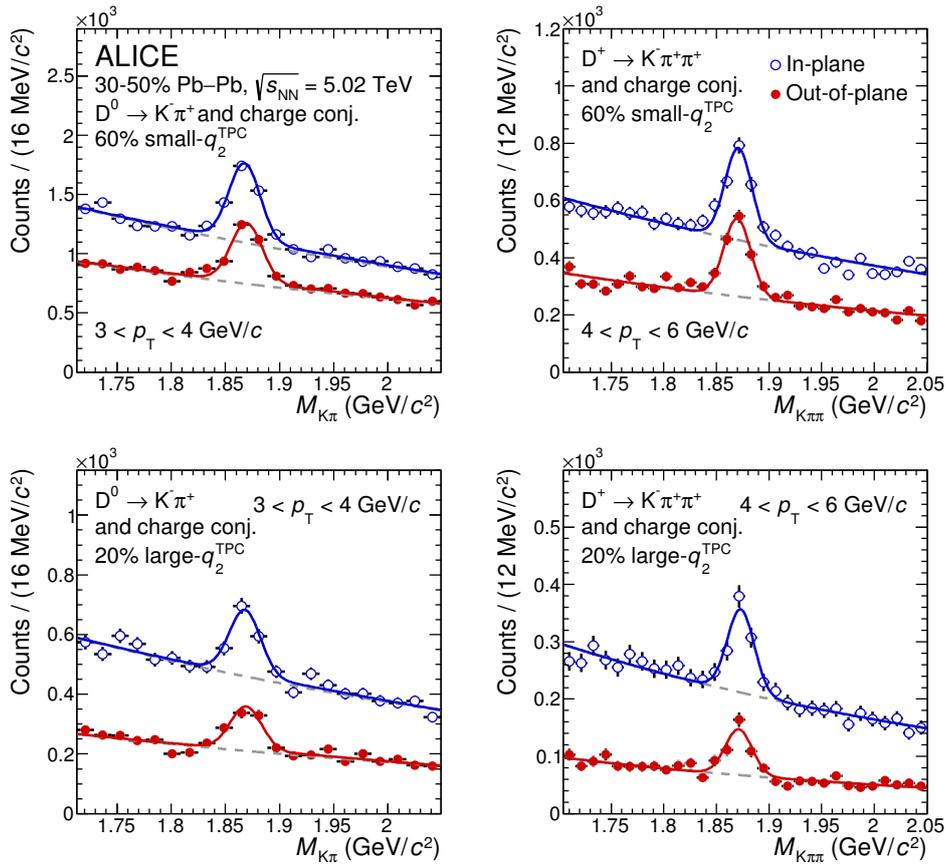}
			\caption{Invariant-mass distributions of $\Dzero$ candidates (left column) and $\Dplus$ candidates (right column) in two $\pt$ intervals for the two $\Delta\varphi$ regions used in the EP method for the 30--50\% $\PbPb$ collisions at $\sqrtsNN=5.02~\TeV$. The top row shows the distributions for the small-$\qTPC$ sample, while the bottom row for the large-$\qTPC$ sample (see text for details). The solid curves represent the total fit functions and the dotted curves the background functions, as described in the text.}
			\label{fig:massplots_ESE} 
			\end{center}
			\end{figure*}

		The in-plane and out-of-plane raw yields were obtained by fitting the invariant-mass distributions $M({\rm K}\pi)$ for $\Dzero$ candidates, $M({\rm K}\pi\pi)$ for $\Dplus$ candidates and the mass-difference $\Delta M = M({\rm K}\pi\pi)-M({\rm K}\pi)$ distributions for $\Dstar$ candidates in each centrality class. The fit function was composed by a Gaussian distribution to describe the signal and an exponential term for the background of $\Dzero$ and $\Dplus$ candidates or by a threshold function multiplied by an exponential function, $a\sqrt{\Delta m - m_{\pi}}\cdot e^{b(\Delta m - m_{\pi})} $, for the $\Dstar$ background. Since the invariant-mass resolution does not exhibit any dependence on $\Delta\varphi$ or $q_2$, the width of the Gaussian, for each D-meson species and $\pt$ interval, was fixed to that obtained from a fit to the invariant-mass distribution integrated over $\Delta\varphi$ and $q_2$, where the signal has higher statistical significance. In addition, for the determination of the $\Dzero$-meson yield, the contribution of signal candidates present in the invariant-mass distribution with the wrong K-$\pi$ mass assignment was taken into account by including an additional term in the fit function, parametrised with a double-Gaussian shape~\cite{Abelev:2014ipa} determined with Monte Carlo simulations. The contribution of the reflected signal, 2--5\% under the $\Dzero$-peak region depending on $\pt$, was considered as background and therefore not included in the raw yield. Examples of invariant-mass fits for the three D-meson species in the unbiased sample in the 10--30\% centrality class and for $\Dzero$ and $\Dplus$ mesons in the ESE-selected samples in the 30--50\% centrality class are shown in Fig.~\ref{fig:massplots_unb} and Fig.~\ref{fig:massplots_ESE}, respectively.
			
			The measured raw D-meson yields contain a feed-down contribution which, depending on the D-meson species, $\pt$ and the topological selections, can vary between 5\% and 20\%. The strategy adopted to correct the observed $v_2$ for the fraction of prompt D mesons in the measured raw yields is the same as the one used in~\cite{Acharya:2017qps}, and it is described in the following. The observed $v_2$ can be expressed as a linear combination of the prompt (D mesons coming directly from the hadronisation of a c-quark or from the decay of an excited open charm or charmonium state) and the feed-down contributions
			\begin{equation}
			v_2^{\rm obs} = f_{\rm prompt}v_2^{\rm prompt}+(1-f_{\rm prompt})v_2^{\rm feed\textnormal{-}down},	
			\end{equation}
			where $f_{\rm prompt}$ is the fraction of promptly produced D mesons estimated as a function of $\pt$ with the same method used in~\cite{Acharya:2018hre}. In particular, it is computed using (i) FONLL calculations~\cite{Cacciari:1998it,Cacciari:2001td} for the production cross-section of beauty hadrons, (ii) the beauty-hadron decay kinematics from the EvtGen package~\cite{Lange:2001uf}, (iii) the product of efficiency and acceptance $({\rm Acc}\times\varepsilon)$ from Monte Carlo simulations and (iv) an hypothesis on the nuclear modification factor of feed-down D mesons. The nuclear modification factor is defined as $\Raa=({\rm d}N_{\rm AA}/{\rm d}\pt)/(\langle T_{\rm AA}\rangle{\rm d}\sigma_{\rm pp}/{\rm d}\pt)$, where ${\rm d}N_{\rm AA}/{\rm d}\pt$ and ${\rm d}\sigma_{\rm pp}/{\rm d}\pt$ are the $\pt$-differential yield and production cross section of D mesons in nucleus--nucleus (AA) and pp collisions, respectively, and $\langle T_{\rm AA}\rangle$ is the average nuclear overlap function in the considered centrality class~\cite{Abelev:2013qoq}. 
			The hypothesis $\Raa^{\rm feed\textnormal{-}down}=2\Raa^{\rm prompt}$ was used to estimate the central value of $f_{\rm prompt}$. This choice is motivated by the comparison of the $\Raa$ of prompt D mesons at $\sqrtsNN = 2.76~\TeV$~\cite{Adam:2015nna} with that of J/$\psi$ from beauty-hadron decays at the same energy measured by the CMS Collaboration~\cite{Khachatryan:2016ypw}, which indicates that the charm-hadron production yield is more suppressed than that of the beauty hadrons by about a factor of two. This difference is described by model calculations with parton-mass-dependent energy loss~\cite{Djordjevic:2015hra}.
			The selection efficiency and therefore $f_{\rm prompt}$ are different in the 10--30\% and 30--50\% centrality classes, because of the different geometrical selections applied on the displaced decay-vertex topology. In the case of the ESE selection, the $({\rm Acc}\times\varepsilon)$ is the same for the large-$q_2$ and small-$q_2$ samples, because the same selection criteria were used in the two ESE-selected classes and the efficiency was found not to depend on local particle density. 
			Therefore, considering also the same $\Raa^{\rm feed\textnormal{-}down}$ hypothesis, $f_{\rm prompt}$ resulted to be equal for the two ESE-selected classes and the unbiased sample in the same centrality interval. The uncertainties arising from the FONLL calculation, as well as the variation of the hypothesis on the $\Raa^{\rm feed\textnormal{-}down}$ in the interval $1<\Raa^{\rm feed\textnormal{-}down}/\Raa^{\rm prompt}<3$, were taken into account as systematic uncertainties. The range of variation of $\Raa^{\rm feed\textnormal{-}down}/\Raa^{\rm prompt}$ takes into account the data uncertainties and model variations. The elliptic flow of promptly produced D mesons was obtained assuming $v_2^{\rm feed\textnormal{-}down}=v_2^{\rm prompt}/2$ and considering a flat probability distribution of $v_2^{\rm feed\textnormal{-}down}$ in the interval $[0,v_2^{\rm prompt}]$. This hypothesis was suggested by the measurement of the non-prompt J/$\psi$ $v_2$ in Pb--Pb collisions at $\sqrtsNN=2.76~\TeV$ performed by the CMS Collaboration~\cite{Khachatryan:2016ypw} and by the available models~\cite{Uphoff:2012gb,Aichelin:2012ww,Greco:2007sz}, that indicate $0<v_2^{\rm feed\textnormal{-}down}<v_2^{\rm prompt}$. As a consequence, the systematic uncertainty on $v_2^{\rm prompt}$ related to the feed-down subtraction is estimated by varying the central value of $v_2^{\rm feed\textnormal{-}down}$ by $\pm v_2^{\rm prompt}/\sqrt{12}$, corresponding to $\pm1$ standard deviation of the assumed uniform distribution.
			
\subsection{Non-flow contamination and $\pmb{q_2}$ selectivity}
\label{subsec:nonflw_q2selectivity}
\begin{figure*}[!t]
\begin{center}
\includegraphics[width=1.\textwidth]{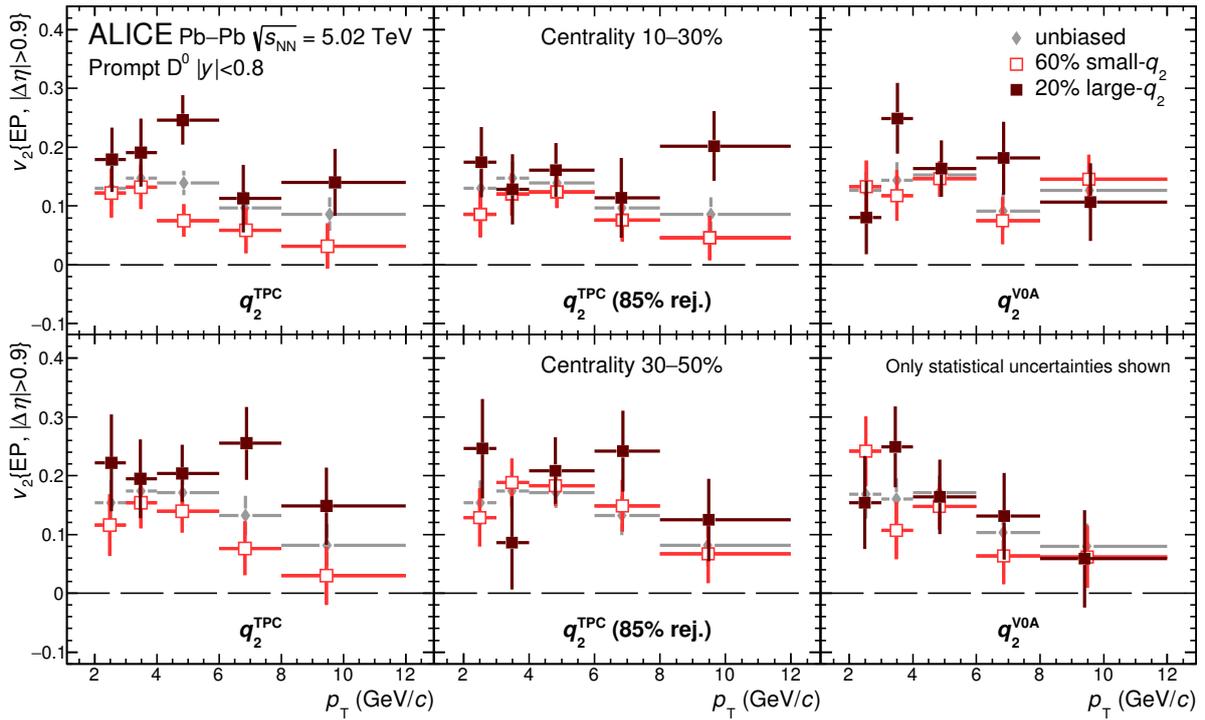}
\caption{Comparison between the $\Dzero$ $v_2$ values measured in the unbiased sample and in the two event-shape classes obtained using TPC and V0A to compute $q_2$, for the 10--30\% (top row) and 30--50\% (bottom row) centrality classes. Only statistical uncertainties are shown.}
\label{fig:Dzerov2ESE_diffq2} 
\end{center}
\end{figure*}

The possible effect of non-flow correlations between the D mesons and the charged particles used in the $q_2$ determination was investigated by comparing the $v_2$ values obtained with the ESE selection based on $\qTPC$ to that obtained by selecting the events according to $\qVZEROA$.
A difference in the results obtained using $\qTPC$ and $\qVZEROA$ can be attributed to different contributions of non-flow correlations, but also to the different eccentricity discriminating power of $q_2$ measured with the two detectors. This discriminating power depends on the magnitude of the elliptic flow, on the multiplicity used in the $q_2$ calculation and on the performance of the detector (i.e. the angular resolution or the linearity of the response as a function of charged-particle multiplicity). To disentangle the two effects, the selectivity of $\qTPC$ was artificially reduced by rejecting randomly 85\% of tracks used for the calculation of $\qTPC$. A similar strategy was used in~\cite{Adam:2015eta}. Figure~\ref{fig:Dzerov2ESE_diffq2} illustrates the comparison of the effect of the ESE selection on the $\Dzero$-meson $v_2$ obtained using $\qTPC$ (left-hand panels), $\qTPC$ with random rejection of 85\% of the tracks (middle panels) and $\qVZEROA$ (right-hand panels), for the 10--30\% (top panels) and the 30--50\% (bottom panels) centrality classes.
The separation between the measurements in the ESE-selected sample with respect to the unbiased one is reduced in the case of the random rejection of the tracks with respect to the default $\qTPC$, confirming the reduced $\qTPC$ selectivity. The results obtained with $\qVZEROA$ are similar to those obtained reducing artificially the selectivity of $\qTPC$, although they are compatible within uncertainties with both $\qTPC$ measurements. This indicates that the statistical precision of the measurement is not sufficient to draw a firm conclusion about non-flow contaminations in the measurement performed by selecting the events according to $\qTPC$. The $\qTPC$-based selection was thus chosen for the evaluation of the results presented in the following sections, except for the comparison of the effect of the ESE selection on the D-mesons and the charged-particle $v_2$, for which the $\qVZEROA$-based selection was used. 

\section{Systematic uncertainties}
\label{sec:systematics}

The values of $v_2$ are affected by systematic 
uncertainties related to (i) the signal extraction from the invariant-mass 
distributions, (ii) the correction for the beauty feed-down contribution, 
(iii) the presence of non-flow effects, and (iv) the centrality dependence of 
the event-plane resolution correction $R_2$.

The uncertainty on the D-meson raw yield extraction from the invariant-mass
distributions of candidates in the in-plane and out-of-plane azimuthal angle
intervals was estimated with a multi-trial approach by repeating the fits several times with different configurations.
In particular, the lower and upper limits of the fit range and the background 
fit function were varied, while the Gaussian width was kept fixed to the
one extracted from the fits to the invariant mass distributions integrated over
$q_2$ and $\Delta\varphi$.
For each fit configuration, the D-meson $v_2$ was calculated from the
in-plane and out-of-plane yields.
The absolute systematic uncertainties were assigned as the r.m.s.\ of the $v_2$ 
distribution resulting from the different fits.
They range from 0.005 to 0.040 in the 30--50\% centrality class and 
from 0.008 to 0.040 in the 10--30\% centrality class, depending on the
 $\pt$ interval and the D-meson species.
Further checks on the stability of the results were performed by 
repeating the procedure leaving the Gaussian width as a free parameter in the 
fits and by using a bin-counting method for the definition of the raw yield.
With the latter method, the signal yield was obtained by counting the 
histogram entries in the invariant-mass region of the signal ($|M-M_{\rm peak}|<3.5\sigma$), after 
subtracting the background contribution estimated from a fit to the side bands ($|M-M_{\rm peak}|>4\sigma$). 
The $v_2$ values obtained from these checks were found to be within the 
uncertainty estimated by varying the fit conditions and therefore
no additional systematic uncertainty was assigned.
For the analysis with ESE selection, further studies were carried out 
by comparing the output of the multiple-trial fit procedure described above 
in the small-$q_2$, large-$q_2$ and $q_2$-integrated samples for each of the 
tested fit configurations.
These checks indicated that this contribution to the systematic uncertainty is 
uncorrelated between the event samples selected based on the $q_2$ 
value.

The contribution of the beauty feed-down correction to the systematic 
uncertainty was estimated varying (i) the quark mass and the 
renormalisation and factorisation scales in the FONLL calculations; (ii) the
$R_{\rm AA}^{\rm feed\textnormal{-}down}$ hypothesis; and (iii) the $v_{2}^{\rm feed\textnormal{-}down}$ hypothesis as described in Sec.~\ref{sec:analysis}.
The value of the corresponding absolute systematic uncertainty ranges from 0.001 to 0.030 depending on the D-meson species and $\pt$ as well as on the ESE-selected class.

The systematic uncertainty on the event-plane resolution correction factor 
$R_2$ has two contributions, which are common to the unbiased, small-$q_2$ and 
large-$q_2$ samples.
The first one originates from possible non-flow effects affecting the 
estimation of $R_2$, when the particles reconstructed in the two semivolumes of the TPC are used as sub-events.
It was estimated by comparing the value of $R_2$ obtained by introducing 
two different pseudorapidity gaps ($\Delta\eta=0.2$ and $\Delta\eta=0.4$)
between the sub-events of the TPC tracks with positive/negative $\eta$.
The second contribution is due to the centrality dependence of $R_2$ within
the classes used in the analysis.
The central value of $R_2$ was computed from the three sub-event correlations 
averaged over the events in the 10--30\% and 30--50\% intervals.
The uncertainty was estimated by comparing this value with those 
obtained as weighted averages of the $R_2$ values in narrow centrality 
intervals, using as weights either the D-meson yields or the average number of 
nucleon--nucleon collisions. 
A systematic uncertainty of 2\% on $R_2$ was assigned based on these 
studies for all centrality and ESE-selected classes.

For the ESE-selected samples, an additional bias on the resolution correction 
factor can originate from autocorrelations because of the usage of TPC tracks (V0A signals)
for both $\qTPC$ ($\qVZEROA$) and $R_2$ determination.
In particular, the selection on $\qTPC$ can bias the correlation between the 
sub-events of charged particles reconstructed in the TPC with 
$0<\eta<0.8$ and with $-0.8<\eta<0$ used in the three sub-event 
calculation of $R_2$.
To estimate this systematic uncertainty, an alternative approach to compute
$R_2$ was utilised, which is based on (i) the correlations between the 
sub-events reconstructed with the V0 and half of the TPC tracks 
(with $\eta<0$) and (ii) the assumption that the ratio of the variables  
$\chi_{\rm V0}$ and $\chi_{{\rm TPC},\eta<0}$ governing the event plane 
resolution (see Ref.~\cite{Poskanzer:1998yz} for its definition) 
is the same in the unbiased and ESE-selected samples.
The difference between the $R_2$ values obtained with this approach and the
three sub-event method, which amounts to 3\% and 5\% in the 
10--30\% and 30--50\% centrality classes, respectively, was assigned as systematic uncertainty on the ESE-selected samples. The same procedure was adopted for the samples selected using the $\qVZEROA$. In this case, the systematic uncertainty was estimated to be of the order of 1\% for the large-$\qVZEROA$ sample, while negligible for the small-$\qVZEROA$ sample, for both the centrality classes.

As discussed in Sec.~\ref{subsec:nonflw_q2selectivity}, a further bias in the analyses with $\qTPC$-based selection could be induced by non-flow 
correlations between the D meson and the sample of tracks used for the $q_2$ 
measurement, which can include charged particles originating from the 
fragmentation of the charm quarks.
To further study this effect, the analysis with $\qTPC$-based selection was repeated introducing 
a ``jet-veto'' pseudorapidity gap of $|\Delta\eta|=0.1$ units 
between each D-meson candidate and the tracks used to measure $\qTPC$.
Since no significant difference was observed, no systematic uncertainty was 
assigned.
\section{Results}
\begin{figure*}[!t]
\begin{center}
\includegraphics[width=1.\textwidth]{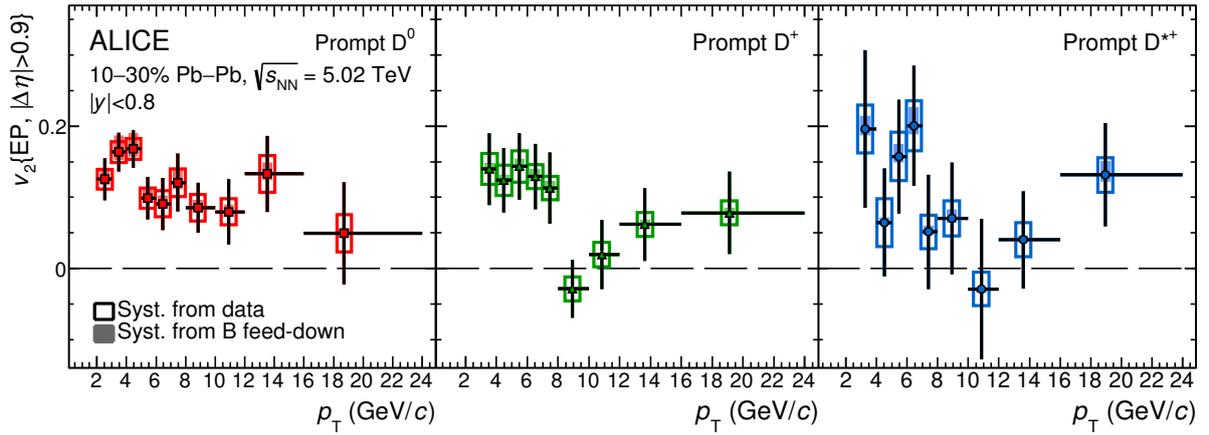}
\caption{Elliptic flow coefficient $v_2$ as a function of $\pt$ for prompt $\Dzero$, $\Dplus$, and $\Dstar$ mesons for $\PbPb$ collisions at $\sqrtsNN=5.02~\TeV$ in the centrality class 10--30\%. The symbols are positioned horizontally at the average $\pt$ of the reconstructed D mesons. Vertical error bars represent the statistical uncertainty, empty boxes the systematic uncertainty associated with the D-meson anisotropy measurement and the event-plane resolution. Shaded boxes show the uncertainty due to the feed-down from beauty-hadron decays.}
\label{fig:DzDpDs_V2_1030} 
\end{center}
\end{figure*}

In Fig.~\ref{fig:DzDpDs_V2_1030} the elliptic flow coefficient $v_2$ of prompt $\Dzero$, $\Dplus$, and $\Dstar$ mesons is reported as a function of $\pt$ in the centrality class 10--30\%. The symbols are positioned at the average $\pt$ of the reconstructed D mesons, which is determined as the average of the $\pt$ distribution of candidates in the signal invariant-mass region, after subtracting the contribution of the background candidates estimated from the side bands. The systematic uncertainty of the feed-down correction is displayed separately in the figure. The $v_2$ of $\Dzero$, $\Dplus$ and $\Dstar$ mesons is consistent among the various species and larger than zero in the interval $2<\pt<8~\GeV/c$. 

\begin{figure*}[!b]
\begin{center}
\includegraphics[width=.9\textwidth]{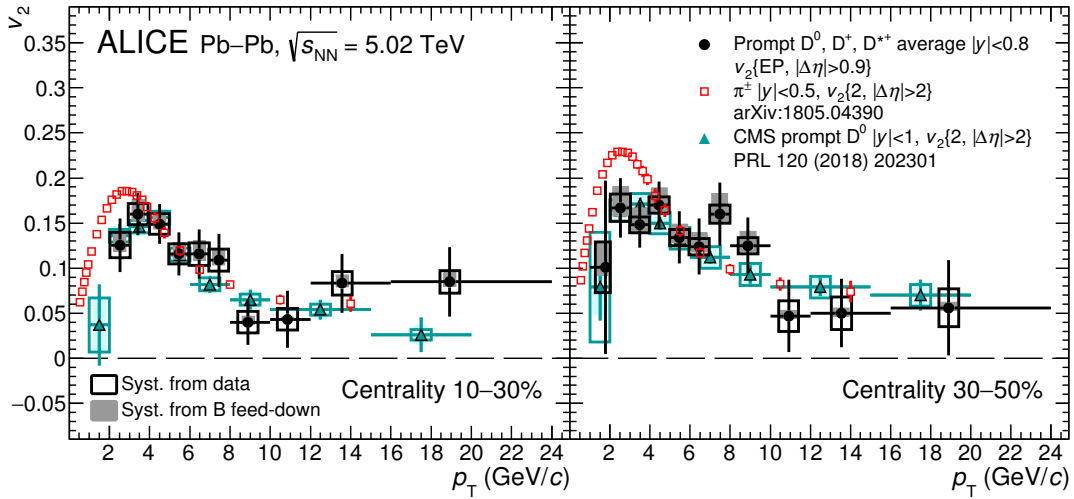}
\caption{Average $\Dzero$, $\Dplus$, $\Dstar$ $v_2$ as a function of $\pt$ for $\PbPb$ collisions at $\sqrtsNN=5.02~\TeV$ in the centrality classes 10--30\% (left) and 30--50\%~\cite{Acharya:2017qps} (right). The comparison with the measurement of the $\Dzero$ $v_2$ by the CMS Collaboration~\cite{Sirunyan:2017plt} and the charged pion $v_2$~\cite{Acharya:2018zuq} in the same centrality intervals is also shown.}
\label{fig:Daverage_V2_1030_3050} 
\end{center}
\end{figure*}

The average $v_2$ and $\pt$ of prompt $\Dzero$, $\Dplus$, $\Dstar$ mesons as a function of $\pt$ was computed by using the inverse of the squared absolute statistical uncertainties as weights and is reported in the left panel of Fig.~\ref{fig:Daverage_V2_1030_3050}. 
The systematic uncertainties were propagated by considering the contribution from the event-plane resolution $R_2$ and the feed-down correction as correlated among the D-meson species. In the right panel of Fig.~\ref{fig:Daverage_V2_1030_3050}, the average $v_2$ of $\Dzero$, $\Dplus$, and $\Dstar$ as a function of $\pt$ in the centrality class 30--50\% taken from~\cite{Acharya:2017qps} is reported. The measurements in both centrality classes are compatible within uncertainties with the $\Dzero$-meson $v_2$ measured with the Scalar Product (SP) method~\cite{Adler:2002pu,Voloshin:2008dg} by the CMS Collaboration in $|y|<1$~\cite{Sirunyan:2017plt}. The charged-pion $v_2$ measured in $|y|<0.5$ by the ALICE Collaboration using the SP method~\cite{Acharya:2018zuq} is also superimposed for comparison. The D-meson $v_2$ is similar in magnitude to that of $\pi^\pm$ for $4<\pt<10~\GeV/c$. In the region $\pt<4~\GeV/c$, where a mass ordering for light hadrons is observed and described by hydrodynamical calculations~\cite{Abelev:2014pua}, the values of the D-meson $v_2$ are slightly lower than those of $\pi^\pm$, but compatible within uncertainties.  

\begin{figure*}[!t]
\begin{center}
\includegraphics[width=.9\textwidth]{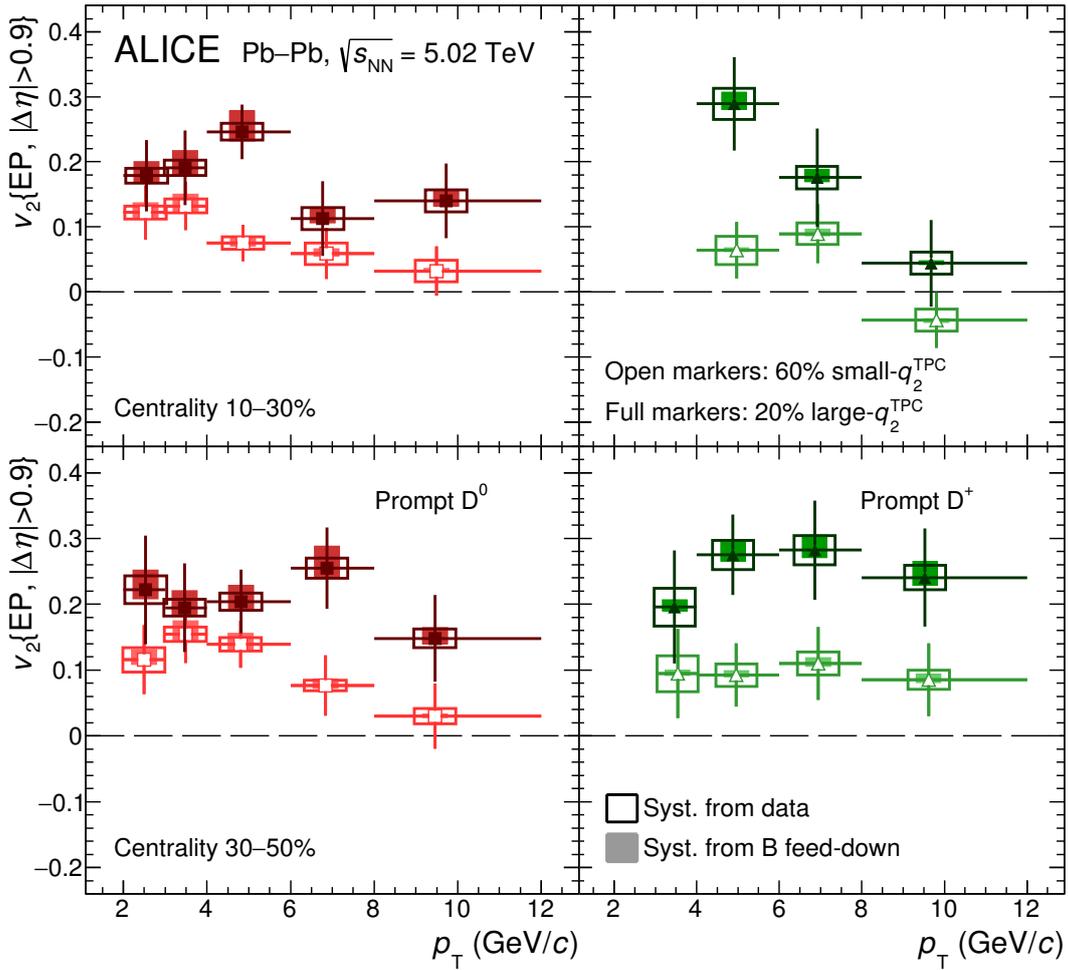}
\caption{$\Dzero$ (left column) and $\Dplus$ (right column) $v_2$ as a function of $\pt$ for the small-$\qTPC$ and large-$\qTPC$ samples (see text for details), in $\PbPb$ collisions at $\sqrtsNN=5.02~\TeV$ in the 10--30\% (top row) and 30--50\% (bottom row) centrality classes. The symbols are positioned horizontally at the average $\pt$ of the reconstructed D mesons. Vertical error bars represent the statistical uncertainty, empty boxes the systematic uncertainty associated with the D-meson anisotropy measurement and the event-plane resolution. Shaded boxes show the uncertainty due to feed-down from beauty-hadron decays.} 
\label{fig:DzDp_ESEV2_1030_3050} 
\end{center}
\end{figure*}
Figure \ref{fig:DzDp_ESEV2_1030_3050} shows the prompt $\Dzero$ and $\Dplus$ $v_2$ as a function of $\pt$ in the small-$\qTPC$ and large-$\qTPC$ samples, in the centrality classes 10--30\% (top row) and 30--50\% (bottom row). The measurement of the $\Dstar$ $v_2$ in the ESE-selected samples was not possible due to the small statistical significance, while the measurements of $\Dzero$ and $\Dplus$ mesons were performed in wider $\pt$ intervals compared to the unbiased $v_2$ measurement and in the reduced range $2<\pt<12~\GeV/c$, due to the limited size of the data sample. The measurements of the $v_2$ of the two different D-meson species in the ESE-selected classes are compatible with each other within uncertainties. Also in this case, the symbols are positioned at the average D-meson $\pt$ determined as described above. 
\begin{figure*}[!t]
\begin{center}
\includegraphics[width=.9\textwidth]{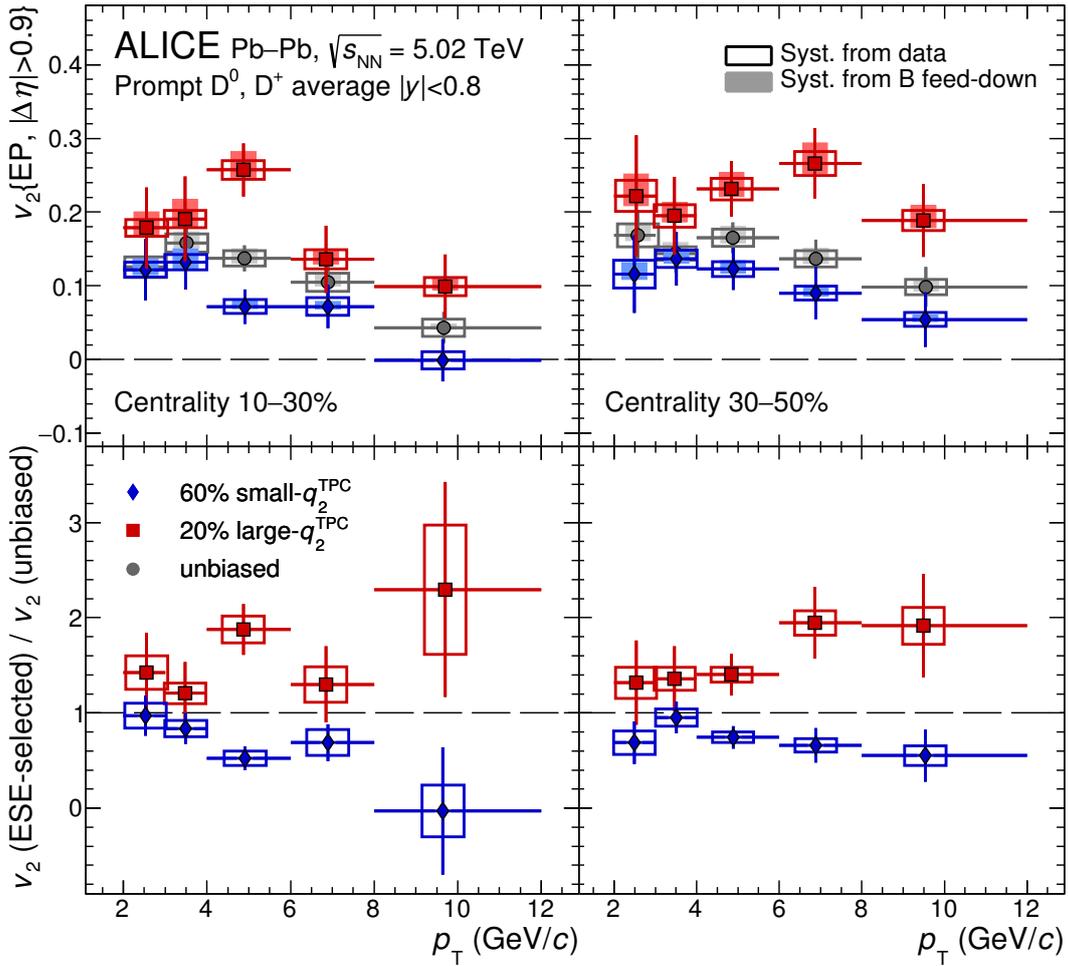}
\caption{Top panels: average of $\Dzero$ and $\Dplus$ $v_2$ as a function of $\pt$ for $\PbPb$ collisions at $\sqrtsNN=5.02~\TeV$ in the small-$\qTPC$, large-$\qTPC$ (see text for details) and unbiased samples, in the 10--30\% (left) and 30--50\% (right) centrality classes. Bottom panels: ratios of the measured $v_2$ in the ESE-selected classes to the one obtained from the unbiased sample.}
\label{fig:Daverage_ESEV2_1030_3050} 
\end{center}
\end{figure*}

The average $v_2$ of $\Dzero$ and $\Dplus$ mesons has been calculated in the small-$\qTPC$ and large-$\qTPC$ samples with the same weighted average procedure described above. It is shown for the two considered centrality classes in the top panels of Fig.~\ref{fig:Daverage_ESEV2_1030_3050} together with the $v_2$ measured in the unbiased sample, recalculated in the same $\pt$ intervals of the ESE analysis. In the bottom panels of the same figure, the ratio of the average D-meson $v_2$ from the ESE-selected samples with respect to that of the unbiased samples is illustrated. The statistical uncertainties on the ratio were propagated taking into account the degree of correlation between the measured yields in the small-$\qTPC$ (large-$\qTPC$) and the unbiased sample. The systematic uncertainties were propagated considering the contribution from the centrality dependence and the non-flow contaminations among sub-events of $R_2$ as well as the feed-down correction as correlated between the measurements in the ESE-selected and the unbiased samples. 

The observation of a flat ratio as a function of $\pt$ for light hadron $v_2$ with ESE-selection at $\sqrtsNN = 2.76~\TeV$ indicated that the $q_2$ value is connected to a global property of the event~\cite{Adam:2015eta}. 
For D mesons, the modification of the $v_2$ in the $\qTPC$-selected samples is compatible within uncertainties with a flat behaviour as a function of $\pt$ for both the 10--30\% and the 30--50\% centrality classes. However, the current precision of the measurement does not allow to exclude a $\pt$ dependence which would indicate the presence of non-flow contaminations. 

\begin{figure*}[!t]
\begin{center}
\includegraphics[width=.9\textwidth]{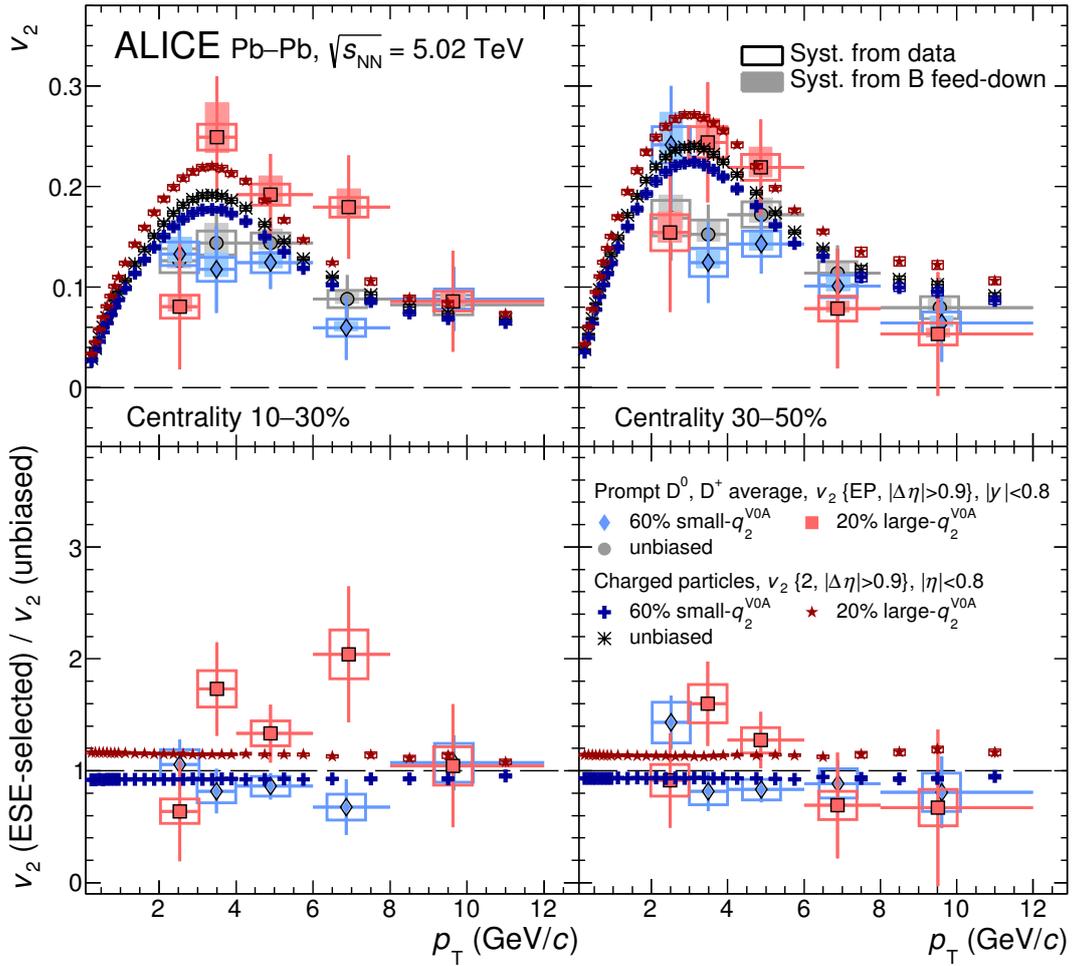}
\caption{Top panels: average of $\Dzero$ and $\Dplus$ $v_2$ as a function of $\pt$ for $\PbPb$ collisions at $\sqrtsNN=5.02~\TeV$ in the small-$\qVZEROA$, large-$\qVZEROA$ (see text for details) and unbiased samples, in the 10--30\% (left) and 30--50\% (right) centrality classes. The charged-particle $v_2$ obtained at the same energy, centrality classes and ESE samples are superimposed for comparison. Bottom panels: ratios of the measured $v_2$ in the ESE-selected classes to the one obtained from the unbiased sample.}
\label{fig:Daverage_ESEV2ratios_q2V0A_1030_3050} 
\end{center}
\end{figure*}

Selecting the 20\% (60\%) highest (lowest) $\qTPC$ sample leads to a change of about 40\% (25\%) in the measured $v_2$. The corresponding variation of the average $\qTPC$ in the ESE-selected classes was found to be about 65\% and 30\% in the large-$\qTPC$ and small-$\qTPC$ samples, respectively. The increase (decrease) of the D-meson $v_2$ and the average $\qTPC$ observed in the large-$\qTPC$ (small-$\qTPC$) sample with respect to the unbiased one is similar within uncertainties in the two centrality intervals considered.
	Considering as null hypothesis $v_2({\rm large}\textnormal{-}\qTPC) = v_2({\rm small}\textnormal{-}\qTPC)$, the probability to observe the measured positive $\Delta v_2  = v_2({\rm large}\textnormal{-}\qTPC) - v_2({\rm small}\textnormal{-}\qTPC)$ in the full $\pt$ range of the measurement, corresponds to a significance of about $4\sigma$, taking into account both statistical and systematic uncertainties in each centrality class. It is however important to keep in mind that part of the observed effect could be slightly enlarged by non-flow contaminations, as previously mentioned.
\begin{figure*}[!t]
\begin{center}
\includegraphics[width=.9\textwidth]{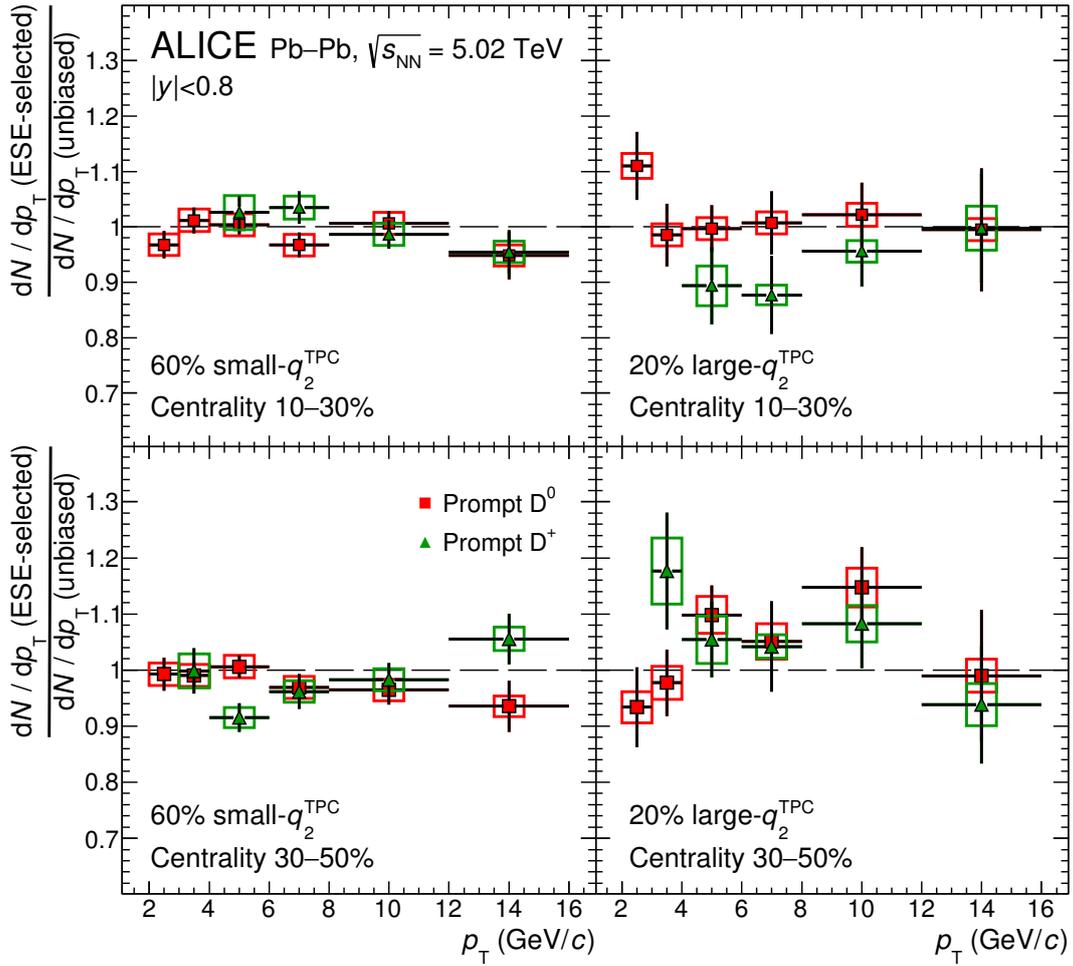}
\caption{Ratio of the yields of $\Dzero$ and $\Dplus$ mesons measured as a function of $\pt$ in the small-$\qTPC$ (left column) and large-$\qTPC$ (right column) samples (see text for details) to that in the unbiased sample, in $\PbPb$ collisions at $\sqrtsNN=5.02~\TeV$ for the 10--30\% (top row) and 30--50\% (bottom row) centrality classes. Vertical error bars represent the statistical uncertainty, empty boxes the total systematic uncertainty.}
\label{fig:DzDp_ESEyieldratios_1030_3050} 
\end{center}
\end{figure*}

The effect of the ESE selection on the D-meson $v_2$ was compared to that observed for charged particles. For this comparison, the ESE selection was performed using $\qVZEROA$, in order to avoid autocorrelations and non-flow contaminations. In the top panels of Fig.~\ref{fig:Daverage_ESEV2ratios_q2V0A_1030_3050}, the average $\Dzero$ and $\Dplus$ $v_2$ in the ESE-selected and unbiased samples in the 10--30\% (left panel) and 30--50\% (right panel) centrality classes are depicted together with the charged-particle $v_2$ measured at the same energy, centrality classes and ESE-selected samples. The charged-particle $v_2$ was measured with the SP method considering reconstructed tracks with $|\eta|<0.8$ and $0.2<\pt<12~\GeV/c$, selected as in Ref.~\cite{Acharya:2018zuq}. 
The bottom panels of the same figure show the ratios of the $v_2$ measured in the ESE-selected samples with respect to the unbiased one. The ratios between the charged-particle $v_2$ show almost no $\pt$ dependence, confirming that the usage of the $\qVZEROA$ provides a selection of a global property of the collision. The relative variation of the charged-particle $v_2$ in the large-$\qVZEROA$ and small-$\qVZEROA$ samples was found to be of about 14-15\% and 7-8\%, respectively. These values reflect the reduced sensitivity of the ESE selection obtained using the V0A with respect to that based on TPC tracks. The ratios of the average D-meson $v_2$ in the ESE-selected samples with respect to the unbiased one were found to be compatible within uncertainties with those of charged particles in the corresponding samples, suggesting that the response to the ESE selection is similar for D mesons and the bulk of light hadrons. However, given the reduced selectivity of $\qVZEROA$ and the current experimental uncertainties, the ratios of the average D-meson $v_2$ are also compatible with unity, and therefore a firm conclusion cannot be drawn. Nevertheless, the comparison between D mesons and charged particles will be crucial for future larger data samples, to better asses the magnitude of the correlation between the D-meson and the soft-hadron $v_2$.

To study a possible interplay between the azimuthal anisotropy of the event and the charm-quark radial flow (at low/intermediate $\pt$) and in-medium energy loss (at high $\pt$), the yields of prompt $\Dzero$ and $\Dplus$ mesons have been measured in six transverse momentum intervals in the range $2<\pt<16~\GeV/c$, in the small-$\qTPC$ and large-$\qTPC$ samples. 

The D-meson raw yields integrated over $\Delta\varphi$ were extracted from the fits to the invariant-mass distributions in the ESE-selected and unbiased classes and normalised to the corresponding number of events in the considered sample. As described in Sec.~\ref{sec:analysis}, the selection and reconstruction efficiencies of prompt D mesons do not show any dependence on $q_2$ within the ESE selections considered in this analysis, therefore no correction to the raw yields was applied. The fraction of prompt D mesons, $f_{\rm prompt}$, was estimated using the same strategy adopted for the $v_2$ measurement and it is the same in the ESE-selected and the unbiased samples.

The ratio of the D-meson yields in the small-$\qTPC$ (large-$\qTPC$) sample to those in the unbiased sample are shown in Fig.~\ref{fig:DzDp_ESEyieldratios_1030_3050} as a function of $\pt$ in the 10--30\% (top row) and 30--50\% (bottom row) centrality classes. The systematic uncertainty on the raw D-meson yield extraction was evaluated directly on the ratio of the yields, applying the same strategy used for the $v_2$ (see Sec.~\ref{sec:systematics}). The systematic uncertainty on the reconstruction and selection efficiency, arising from a possible imperfect description of the data in the Monte Carlo simulations, cancels out in the ratio, since the efficiency is the same in the two ESE-selected classes. 

\begin{figure*}[!t]
\begin{center}
\includegraphics[width=.9\textwidth]{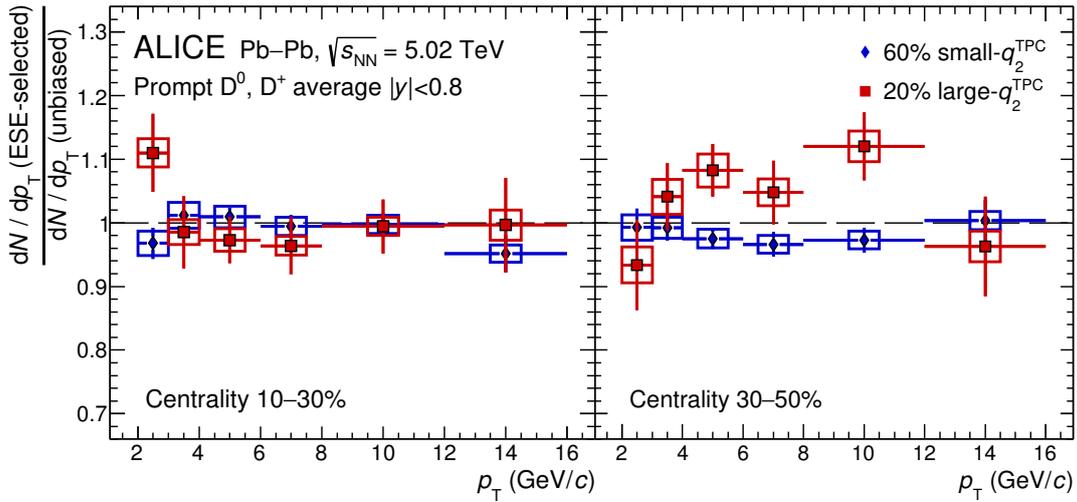}
\caption{Average of the ratio of $\Dzero$ and $\Dplus$ yields measured as a function of $\pt$ in the ESE-selected samples to those in the unbiased sample, in $\PbPb$ collisions at $\sqrtsNN=5.02~\TeV$ for the 10--30\% (left panel) and 30--50\% (right panel) centrality classes. Vertical error bars represent the statistical uncertainty, empty boxes the total systematic uncertainty.}
\label{fig:AverageD_ESEyieldratios_1030_3050} 
\end{center}
\end{figure*}

The average of the ratio of the $\Dzero$ and $\Dplus$ yields in the small-$\qTPC$ (large-$\qTPC$) sample to those in the unbiased sample is depicted in Fig.~\ref{fig:AverageD_ESEyieldratios_1030_3050}. It was computed by using the inverse of the squared relative statistical uncertainties as weights. 

 In the 10--30\% centrality class, the ratio between the D-meson yields in ESE-selected samples to those in the unbiased sample was found to be compatible with unity in the measured $\pt$ range. In the 30--50\% centrality class, the central values of the D-meson per-event yields in the large-$\qTPC$ (small-$\qTPC$) samples were found to be higher (lower) than those in the unbiased sample in all the measured $\pt$ intervals in the range $3<\pt< 12~\GeV/c$. However, the ratios between the yields in the ESE-selected samples to the unbiased yields are compatible with unity within about one standard deviation.

 In the light-hadron sector, the effect induced by the correlation between radial and elliptic flow, attributed to a larger initial density in more anisotropic events, was observed to be of the order of 5\% for charged pions with $\pt\approx4~\GeV/c$ in mid-central $\PbPb$ collisions at $\sqrtsNN=2.76~\TeV$~\cite{Adam:2015eta}. Since the ratio between the D-meson yields in ESE-selected samples to those in the unbiased sample was found to be compatible with unity, a possible similar effect is expected to be smaller than the present experimental uncertainties, which do not allow for any conclusion.

\section{Summary}
The first application of the event-shape engineering technique to the measurement of D-meson production in Pb--Pb collisions at $\sqrtsNN=5.02~\TeV$ has been presented. 

The elliptic flow of $\Dzero$, $\Dplus$, and $\Dstar$ mesons at mid-rapidity in the 10--30\% (30--50\%) centrality class was measured with the event-plane technique and found to be larger than zero in the transverse momentum interval $2<\pt<8(10)~\GeV/c$ and similar in magnitude to that of charged pions for $\pt>4~\GeV/c$, while slightly lower for $\pt<4~\GeV/c$, in the same centrality class. 

The $v_2$ coefficient of $\Dzero$ and $\Dplus$ mesons was measured in events with different magnitude of the average bulk elliptic flow, quantified by the value of $q_2$ measured using TPC tracks to maximise the selectivity. The observation of a larger (smaller) D-meson $v_2$ in events with large-$\qTPC$ (small-$\qTPC$) values confirms a correlation between D-meson azimuthal anisotropy and the collective expansion of the bulk of light hadrons. When using the V0A to measure $q_2$ in order to reduce non-flow contaminations and autocorrelations, the variation of the D-meson $v_2$ in the small-$\qVZEROA$ and large-$\qVZEROA$ samples was found to be compatible within uncertainties with that of charged particles, suggesting a similar response to the ESE selection.  

	The ratio of the $\pt$-differential yields measured in the ESE-selected samples with respect to those in the unbiased sample was found to be compatible with unity in both the 10--30\% and 30--50\% centrality classes, with a possible indication of larger D-meson yield for $3<\pt<12~\GeV/c$ in events with higher-than-average bulk elliptic flow in the 30--50\% centrality class. 
	With the current uncertainties no firm conclusion can be drawn on the possible interplay between the initial spatial anisotropy and the charm-quark energy loss and radial flow.
		
	The measurements presented in this paper open the way to the study of heavy-quark production with the Event-Shape Engineering technique, which offers a new possibility to understand the correlation of heavy-quark and bulk properties. An improved precision is expected to be achieved with future data samples that will be collected in 2018 and during Run 3 and 4 of the LHC~\cite{Abelevetal:2014dna,Abelevetal:2014cna}.

\newenvironment{acknowledgement}{\relax}{\relax}
\begin{acknowledgement}
\section*{Acknowledgements}

The ALICE Collaboration would like to thank all its engineers and technicians for their invaluable contributions to the construction of the experiment and the CERN accelerator teams for the outstanding performance of the LHC complex.
The ALICE Collaboration gratefully acknowledges the resources and support provided by all Grid centres and the Worldwide LHC Computing Grid (WLCG) collaboration.
The ALICE Collaboration acknowledges the following funding agencies for their support in building and running the ALICE detector:
A. I. Alikhanyan National Science Laboratory (Yerevan Physics Institute) Foundation (ANSL), State Committee of Science and World Federation of Scientists (WFS), Armenia;
Austrian Academy of Sciences and Nationalstiftung f\"{u}r Forschung, Technologie und Entwicklung, Austria;
Ministry of Communications and High Technologies, National Nuclear Research Center, Azerbaijan;
Conselho Nacional de Desenvolvimento Cient\'{\i}fico e Tecnol\'{o}gico (CNPq), Universidade Federal do Rio Grande do Sul (UFRGS), Financiadora de Estudos e Projetos (Finep) and Funda\c{c}\~{a}o de Amparo \`{a} Pesquisa do Estado de S\~{a}o Paulo (FAPESP), Brazil;
Ministry of Science \& Technology of China (MSTC), National Natural Science Foundation of China (NSFC) and Ministry of Education of China (MOEC) , China;
Ministry of Science and Education, Croatia;
Centro de Aplicaciones Tecnol\'{o}gicas y Desarrollo Nuclear (CEADEN), Cubaenerg\'{\i}a, Cuba;
Ministry of Education, Youth and Sports of the Czech Republic, Czech Republic;
The Danish Council for Independent Research | Natural Sciences, the Carlsberg Foundation and Danish National Research Foundation (DNRF), Denmark;
Helsinki Institute of Physics (HIP), Finland;
Commissariat \`{a} l'Energie Atomique (CEA) and Institut National de Physique Nucl\'{e}aire et de Physique des Particules (IN2P3) and Centre National de la Recherche Scientifique (CNRS), France;
Bundesministerium f\"{u}r Bildung, Wissenschaft, Forschung und Technologie (BMBF) and GSI Helmholtzzentrum f\"{u}r Schwerionenforschung GmbH, Germany;
General Secretariat for Research and Technology, Ministry of Education, Research and Religions, Greece;
National Research, Development and Innovation Office, Hungary;
Department of Atomic Energy Government of India (DAE), Department of Science and Technology, Government of India (DST), University Grants Commission, Government of India (UGC) and Council of Scientific and Industrial Research (CSIR), India;
Indonesian Institute of Science, Indonesia;
Centro Fermi - Museo Storico della Fisica e Centro Studi e Ricerche Enrico Fermi and Istituto Nazionale di Fisica Nucleare (INFN), Italy;
Institute for Innovative Science and Technology , Nagasaki Institute of Applied Science (IIST), Japan Society for the Promotion of Science (JSPS) KAKENHI and Japanese Ministry of Education, Culture, Sports, Science and Technology (MEXT), Japan;
Consejo Nacional de Ciencia (CONACYT) y Tecnolog\'{i}a, through Fondo de Cooperaci\'{o}n Internacional en Ciencia y Tecnolog\'{i}a (FONCICYT) and Direcci\'{o}n General de Asuntos del Personal Academico (DGAPA), Mexico;
Nederlandse Organisatie voor Wetenschappelijk Onderzoek (NWO), Netherlands;
The Research Council of Norway, Norway;
Commission on Science and Technology for Sustainable Development in the South (COMSATS), Pakistan;
Pontificia Universidad Cat\'{o}lica del Per\'{u}, Peru;
Ministry of Science and Higher Education and National Science Centre, Poland;
Korea Institute of Science and Technology Information and National Research Foundation of Korea (NRF), Republic of Korea;
Ministry of Education and Scientific Research, Institute of Atomic Physics and Romanian National Agency for Science, Technology and Innovation, Romania;
Joint Institute for Nuclear Research (JINR), Ministry of Education and Science of the Russian Federation and National Research Centre Kurchatov Institute, Russia;
Ministry of Education, Science, Research and Sport of the Slovak Republic, Slovakia;
National Research Foundation of South Africa, South Africa;
Swedish Research Council (VR) and Knut \& Alice Wallenberg Foundation (KAW), Sweden;
European Organization for Nuclear Research, Switzerland;
National Science and Technology Development Agency (NSDTA), Suranaree University of Technology (SUT) and Office of the Higher Education Commission under NRU project of Thailand, Thailand;
Turkish Atomic Energy Agency (TAEK), Turkey;
National Academy of  Sciences of Ukraine, Ukraine;
Science and Technology Facilities Council (STFC), United Kingdom;
National Science Foundation of the United States of America (NSF) and United States Department of Energy, Office of Nuclear Physics (DOE NP), United States of America.
\end{acknowledgement}

\bibliographystyle{utphys}   
\bibliography{Dmeson_v2ESE}

\newpage
\appendix
\section{The ALICE Collaboration}
\label{app:collab}
\begingroup
\small
\begin{flushleft}

S.~Acharya$^{\rm 139}$, 
F.T.-.~Acosta$^{\rm 20}$, 
D.~Adamov\'{a}$^{\rm 93}$, 
A.~Adler$^{\rm 74}$, 
J.~Adolfsson$^{\rm 80}$, 
M.M.~Aggarwal$^{\rm 98}$, 
G.~Aglieri Rinella$^{\rm 34}$, 
M.~Agnello$^{\rm 31}$, 
N.~Agrawal$^{\rm 48}$, 
Z.~Ahammed$^{\rm 139}$, 
S.U.~Ahn$^{\rm 76}$
% ...

\bigskip

\bigskip 

\textbf{\Large Affiliation Notes}

\bigskip 

% ...
$^{\rm V}$ Also at: Institute of Theoretical Physics, University of Wroclaw, Poland\\

\bigskip

\bigskip 

\textbf{\Large Collaboration Institutes}

\bigskip 

% ...
$^{20}$ Department of Physics, University of California, Berkeley, California, United States\\
% ...
$^{31}$ Dipartimento DISAT del Politecnico and Sezione INFN, Turin, Italy\\
% ...
$^{34}$ European Organization for Nuclear Research (CERN), Geneva, Switzerland\\
% ...
$^{48}$ Indian Institute of Technology Bombay (IIT), Mumbai, India\\
% ...
$^{74}$ Johann-Wolfgang-Goethe Universit\"{a}t Frankfurt Institut f\"{u}r Informatik, Fachbereich Informatik und Mathematik, Frankfurt, Germany\\
% ...
$^{76}$ Korea Institute of Science and Technology Information, Daejeon, Republic of Korea\\
% ...
$^{80}$ Lund University Department of Physics, Division of Particle Physics, Lund, Sweden\\
% ...
$^{93}$ Nuclear Physics Institute of the Czech Academy of Sciences, \v{R}e\v{z} u Prahy, Czech Republic\\
% ...
$^{98}$ Physics Department, Panjab University, Chandigarh, India\\
% ...
$^{139}$ Variable Energy Cyclotron Centre, Homi Bhabha National Institute, Kolkata, India\\
% ...

\bigskip 

\end{flushleft} 
\end{document}